\documentclass{aa}
 
\usepackage{graphicx}
\usepackage{natbib}
\usepackage{subfig}
\usepackage{epsfig} 
\usepackage{ifpdf} 
\usepackage{fancyvrb}
\usepackage{txfonts}
\usepackage{xcolor}
\usepackage{soul}
\usepackage{mathtools}
\usepackage{datetime}
\begin{document} 

\title{Long-term multi-band photometric monitoring of Mrk 501}

\author{ 
Axel Arbet-Engels \inst{1} \and
Dominik Baack \inst{2} \and
Matteo Balbo \inst{3} \and
Adrian Biland \inst{1} \and
Thomas Bretz \inst{1,5} \and
Jens Buss \inst{2} \and
Daniela Dorner \inst{4} \and
Laura Eisenberger \inst{4} \and
Dominik Elsaesser \inst{2} \and
Dorothee Hildebrand \inst{1} \and
Roman Iotov \inst{4} \and
Adelina Kalenski \inst{4} \and
Karl Mannheim \inst{4} \and
Alison Mitchell \inst{1} \and
Dominik Neise \inst{1} \and
Maximilian Noethe \inst{2} \and
Aleksander Paravac \inst{4} \and
Wolfgang Rhode \inst{2} \and
Bernd Schleicher \inst{4} \and
Vitalii~Sliusar \inst{3}\thanks{Corresponding authors, e-mails:  vitalii.sliusar@unige.ch \& roland.walter@unige.ch} \and
Roland Walter \inst{3}$^\star$
}

\institute{
Institute for Particle Physics and Astrophysics, ETH Zurich, Otto Stern Weg 5, CH-8093 Zurich, Switzerland \and 
Fakult\"at Physik, Technische Universit\"at Dortmund, Otto Hahn Str. 4a, D-44227 Dortmund, Germany \and
Department of Astronomy, University of Geneva, Chemin Pegasi 51, CH-1290 Versoix, Switzerland \and 
Institut f\"ur Theoretische Physik und Astrophysik, Universit\"at W\"urzburg, Emil Fischer Str. 31, 97074 W\"urzburg, Germany \and 
Physikalisches Institut III A, RWTH Aachen University, Otto Blumenthal Str., D-52074 Aachen, Germany
}

\date{Received Month Day, Year; accepted Month Day, Year.}
 
\abstract
   {}
   {
    Radio-to-TeV observations of the bright nearby (z=0.034) blazar Markarian\,501 (Mrk\,501), performed from December 2012 to April 2018, are used to study the emission mechanisms in its relativistic jet.}
   {We examined the multi-wavelength variability and the correlations of the light curves obtained by eight different instruments, including the First G-APD Cherenkov Telescope (FACT), observing Mrk\,501 in very high-energy (VHE) gamma-rays at TeV energies. We identified individual TeV and X-ray flares and found a sub-day lag between variability in these two bands.}
   {Simultaneous TeV and X-ray variations with almost zero lag are consistent with synchrotron self-Compton (SSC) emission, where TeV photons are produced through inverse Compton scattering. The characteristic time interval of 5-25 days between TeV flares is consistent with them being driven by Lense-Thirring precession.}
   {}

\maketitle

\section{Introduction\label{sec:introduction}}

Blazars are active galactic nuclei (AGN) with a relativistic jet pointed towards the observer. These sources emit from the radio to the TeVs, and they are the most populous group of objects detected above 100\,GeV. Their emission is non-thermal and is typically characterised by two broad bumps peaking in the infrared to the X-rays, and in the $\gamma$-rays. The low-energy bump is associated with synchrotron radiation of relativistic electrons. Depending on the peak energy of this component, blazars can be subdivided into low (peaking in the infrared), intermediate, or high (peaking in the X-rays) synchrotron-peaked blazars (LBL, IBL, and HBL respectively). Multi-wavelength variability studies provide important insight into the jet inner structure and emission mechanisms. In addition to the electromagnetic radiation, neutrinos have also been associated with some blazars and were used to probe the radiation processes within the jet \citep{2001APh....15..121M,2001ICRC....3.1153M,2016APh....80..115P, 1993A&A...269...67M}.

Markarian 501 (Mrk\,501) is one of the most frequently studied nearby ($z=0.034$) bright blazars and was discovered to be a TeV source by the Whipple imaging atmospheric Cherenkov telescope \citep{1996ApJ...456L..83Q}. Mrk\,501 has been monitored extensively in the radio \citep{Richards_2011ApJS..194...29R,2010ApJ...723.1150P}, V-band \citep{2009arXiv0912.3621S}, X-rays \citep{2020A&A...637A..86M,abdo_2011ApJ...727..129A}, and $\gamma$-rays \citep{abdo_2011ApJ...727..129A,2015arXiv150202582D,Abeysekara_2017ApJ...841..100A,ahnen_2017A&A...603A..31A}.  Although variability is present in all energy bands, it peaks in the TeV, with a flux varying from 0.3 to 10 times that of the Crab nebula (Crab Unit, CU) \citep{1999A&A...349...11A}. In 2012, Mrk\,501 was simultaneously observed by 25 different instruments during a three-month campaign \citep{2018A&A...620A.181A} and was found in an intermediate state with a TeV flux of about 0.5\,CU that reached 4.9\,CU on June 9. The TeV outburst was also accompanied by X-ray flares that were detected by {\it Swift}/XRT. This campaign reported the hardest VHE spectra ever measured, with a power-law index close to 2. The bumps in spectral energy distribution (SED) peaked at above 5\,keV and 0.4\,TeV, respectively, indicating that the source is an extreme HBL (EHBL). This extreme spectrum could be transient, the usual harder-when-brighter TeV spectrum was not observed either, in contrast to previous multi-wavelength campaigns \citep{2011ApJ...729....2A, 2001ApJ...546..898A, 2007ApJ...669..862A, 2011ApJ...727..129A}.

A one-zone SSC model reasonably explains the overall shape of the SED \citep{2012ApJ...758....2B,2015A&A...573A..50A,2018A&A...620A.181A} and accounts for most of its emission. Mismatches were pointed out during some flares \citep[e.g.][]{2018A&A...620A.181A,ahnen_2017A&A...603A..31A}, and were sometimes solved using two different emitting zones \citep{2012ApJ...758....2B,2018A&A...620A.181A} or hadronic processes \citep{2020MNRAS.492.2261S,shukla_2015ApJ...798....2S}. For example, the flare on June 9, 2012, could not be explained using a one-zone SSC model  \citep{2018A&A...620A.181A}, and a second smaller emission zone, mostly contributing to the X-rays and TeVs, was added. The acceleration mechanism in the SSC framework is also unclear, with authors claiming magnetic reconnections (\cite{2018A&A...620A.181A}) or Fermi first-order acceleration \citep{shukla_2015ApJ...798....2S}. Another approach is to consider a photo-hadronic model \citep{2020MNRAS.492.2261S,shukla_2015ApJ...798....2S}, where Fermi-accelerated protons produce the VHE $\gamma$-rays through the photo-pion process. As the most significant gamma-ray variations occur during the high state of Mrk\,501 \citep{abdo_2011ApJ...736..131A,2017A&A...602A..25Z}, comparing models applied to different source states is also difficult.

In this paper, we report observations that took place between 2012 and 2018 from the radio to the TeV, which were part of a long-term multi-wavelength campaign carried out with the FACT telescope. This paper is structured as follows. In Sect.\,\ref{sec:data} we briefly describe the instruments and data reduction techniques we used to obtain the light curves. In Sect.\,\ref{sec:timing} we report the investigations of the variability, auto- and cross-correlations of the light curves, identification and correlations of individual TeV and X-ray flares, and analysis of the correlations between GeV and radio variations. Section\,\ref{sec:discussion} includes the discussion of the main results, with an emphasis on the likely underlying physical processes. Finally, Sect.\,\ref{sec:conclusions} provides a summary of the results and conclusions.

\section{Data and analysis\label{sec:data}}

\begin{figure}[h!]
\centering
\includegraphics[width=\columnwidth]{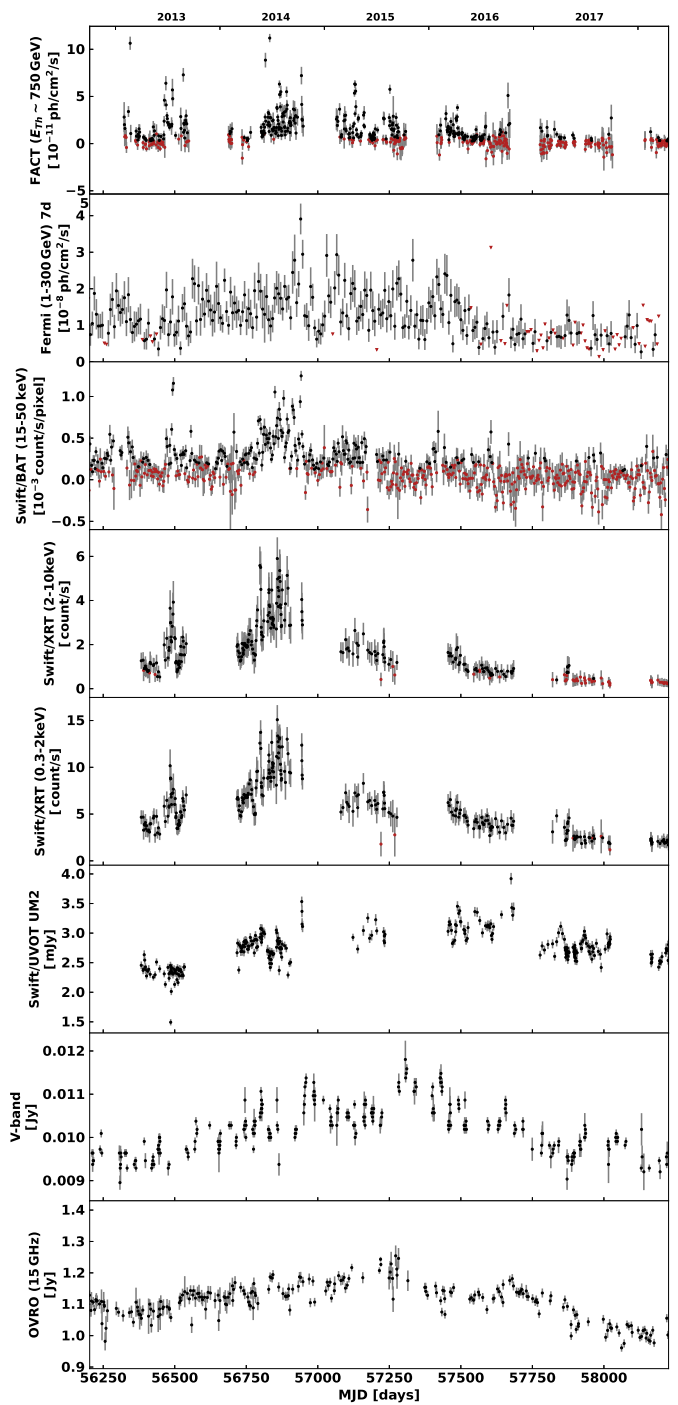}
\caption{Long-term light curves of Mrk\,501 obtained from the radio to the TeVs between December 14, 2012, and April 18, 2018. From top to bottom: FACT ($E_{Th}$ $\sim$750\,GeV) nightly binning, {\it Fermi}-LAT (1-300\,GeV) one-week binning, {\it Swift}/BAT (15-50\,keV) two-day binning, {\it Swift}/XRT (2-10\,keV), {\it Swift}/XRT (0.3-2\,keV), and UM2 {\it Swift}/UVOT V-band optical observations, and radio observations at 15\,GHz. Flux measurement errors are denoted as vertical grey bars. Points with a signal-to-noise ratio lower than 2 are plotted in red (see Sect.~\ref{sec:data} for details about the cleaning procedures). For the {\it Fermi}-LAT light curve, the 95\% flux upper limits (TS<25) are shown with triangles.}
\label{fig:LC}
\end{figure}

To characterise the long-term variability of Mrk\,501, we gathered data taken between December 14, 2012, and April 18, 2018, with eight different instruments observing from the radio to the TeV. 
The instruments and data reduction techniques are described in the following subsections, and the resulting light curves are presented in Fig.~\ref{fig:LC}. The light curves obtained for FACT and {\it Swift}/BAT include negative count rates, as expected when the emission from a source is lower than the sensitivity of a background-dominated instrument.

The {\it Fermi}-LAT light curve is always positive because of the use of a positively defined model in the maximum likelihood fitting. Correlations and variability analyses presented in Sect. \ref{sec:timing} disregard negative or low signal-to-noise ratio (<2 sigma) data points. Flare identification used complete light curves because uncertainties are properly considered in the Bayesian Block algorithm.

\subsection{Radio to GeV observations\label{sec:sub_radio_xrays}}
Mrk\,501 was observed by Owens Valley Radio Observatory (OVRO) 40 meters telescope as part of the {\it Fermi}-LAT high-cadence blazars monitoring program \citep{Richards_2011ApJS..194...29R}. The radio receiver of the OVRO telescope has a centre frequency of 15\,GHz with a 3\,GHz bandwidth and an aperture efficiency $\sim 0.25$. The radio flux densities were obtained using ON-ON technique for dual-beam optics configuration to minimise ground and atmospheric contamination of the signal. The noise at the level of 55\,K, which includes actual thermal noise of 30\,K, the CMB, as well as atmospheric and ground contributions, leads to a $\sim 5\%$ systematic flux uncertainty. Observations were mostly performed twice per week. The data were publicly available online from the Owens Valley Radio Observatory archive\footnote{https://sites.astro.caltech.edu/ovroblazars/data.php. As of submission of this paper, the policy has changed and the data is only available upon request to the \textit{OVRO} collaboration.}. A complete description of the instrument, calibrations, and analysis is provided in \cite{Richards_2011ApJS..194...29R}.

Regular observations of Mrk\,501 were performed in the V-band as part of the optical monitoring of $\gamma$-ray blazars to support {\it Fermi}-LAT observations by the 1.54\,m Kuiper and 2.3\,m\,Bok telescopes at Steward Observatory \citep{2009arXiv0912.3621S}. All the data are publicly available, and for the time period used in this paper, we used data from Cycle 5 to Cycle 10, spanning from September 9, 2012, until March 2018 \footnote{http://james.as.arizona.edu/$\sim$psmith/Fermi/DATA/photdata.html}.

Mrk\,501 observations in three ultraviolet bands (W1, M2, and W2 filters) are available from the {\it Swift} Ultra-Violet and Optical Telescope (UVOT) \citep{roming_2005SSRv..120...95R}. Standard aperture photometry analysis was performed using the {\it Swift}/UVOT software tools from the HEASOFT package (version 6.24). Calibration data were obtained from the CALDB (version 20170922). An aperture of 6 arcsec radius was used for the flux extraction for all the filters. The background flux level was estimated in an annulus with radii 11 or 19 arcsec centred on the location of Mrk\,501. The two regions we used for source and background estimation were verified to not include light from any nearby sources, stray light, or elements of UVOT supporting structures. Dereddening of the fluxed was performed following the prescription from
\cite{2009ApJ...690..163R} using $E(B - V) = 0.0164$ \citep{2011ApJ...737..103S}.

Observations in the X-rays (0.2-10\,keV) were performed by the {\it Swift}/XRT X-ray telescope \citep{2005SSRv..120..165B}. Considering the SED of Mrk\,501, the X-ray light curves from {\it Swift}/XRT were built for the 0.3-2\,keV and 2-10\,keV bands separately to probe the emission process below and above the cutoff energy. {\it Swift}/XRT performed regular high-quality observations of Mrk\,501. The light curve is publicly available using the {\it Swift}-XRT products generation tool \footnote{http://www.{\it Swift}.ac.uk/user\_objects/}. This tool uses \texttt{HEASOFT} software version 6.22, and the complete analysis pipeline is described in \cite{evans_2009MNRAS.397.1177E}.

The Burst Alert Telescope \citep[BAT,][]{0067-0049-209-1-14} on board of the {\it Swift} satellite allows monitoring the complete sky at hard X-rays every few hours. The {\it Swift}/BAT reduction pipeline is described in \cite{2010ApJS..186..378T} and \cite{2013ApJS..207...19B}. Our pipeline is based on the BAT analysis software \texttt{HEASOFT} version 6.13. A first analysis was performed to derive background detector images. We created sky images (task \texttt{batsurvey}) in the eight standard energy bands (in keV: 14 - 20, 20 - 24, 24 - 35, 35 - 50, 50 - 75, 75 - 100, 100 - 150, and 150 - 195) using an input catalogue of 86 bright sources that have the potential of being detected in single pointings. The detector images were then cleaned by removing the contribution of all detected sources (task \texttt{batclean}) and averaged to obtain one background image per day. The variability of the background detector images was smoothed pixel by pixel, fitting the daily background values with a polynomial model. The BAT image analysis was then run using these smoothed averaged background maps. The result of our processing was compared to the standard results presented by the {\it Swift} team (light curves and spectra of bright sources from the {\it Swift}/BAT 70-month survey catalogue\footnote{http://{\it Swift}.gsfc.nasa.gov/results/bs70mon/}), and a very good agreement was found. The {\it Swift}/BAT light curves of Mrk\,501 were built in several energy bands. For each time bin and energy band, a weighted mosaic of the selected data was first produced, and the source region count rate was extracted assuming a fixed source position and shape of the point spread function. The  signal-to-noise ratio of the source varies regularly because of intrinsic variability, its position in the BAT field of view, and the distance to the Sun. The 15-50\,keV two-day bin light curve presented in Fig.~\ref{fig:LC} spans from December 14,  2012, to April  18,  2018, that is, over 29344 orbital periods or almost 5.5 years.

The Large Area Telescope on board the Fermi Gamma-ray Space Telescope (hereafter {\it Fermi}-LAT) is a $\gamma$-ray detector that is sensitive from 20\,MeV to 300\,GeV \citep{2009ApJ...697.1071A, 2009APh....32..193A, 2012ApJS..203....4A, 2012APh....35..346A}. Data are publicly available and have been analysed with the \textit{Fermitools}\footnote{https://fermi.gsfc.nasa.gov/ssc/data/analysis/software/}.

We performed a binned analysis following the {\it Fermi}-LAT team recommendations. We used only photons with energy between 1\,GeV and 300\,GeV within a square of $21^\circ \times 21^\circ$ centred on Mrk\,501, flagged with \texttt{evclass=128} and \texttt{evtype=3}, and with a maximum zenith angle of $90^\circ$. Data were processed with the PASS8\footnote{http://fermi.gsfc.nasa.gov/ssc/data/analysis/documentation/\\/Pass8$\_$usage.html}, using the \textit{P8R3\_SOURCE\_V2} instrument response function. To model the region surrounding Mrk\,501, we used the sources in the 4FGL catalogue \citep{2020ApJS..247...33A}, up to $5^{\circ}$ outside the box, and we described the Galactic diffuse emission and the isotropic extra-Galactic component with \textit{gll\_iem\_v07} and \textit{iso\_P8R3\_SOURCE\_V2\_v1}, respectively\footnote{http://fermi.gsfc.nasa.gov/ssc/data/access/lat/BackgroundModels}. The best fit over the entire time data sample, describing Mrk\,501 with a log-parabola model $dN/dE = N(E/E_{\rm b})^{-\alpha + \beta(Log(E/E_b))}$ , yielded $N=(4.07\pm 0.06)\times 10^{-12}$~$\rm cm^{-2}~s^{-1}~MeV^{-1}$, $\alpha=1.73 \pm 0.01$, $\beta=0.013 \pm 0.005$, with $E_{\rm b}=1476.73$~ MeV. The same fit also yielded a normalisation of $0.97 \pm 0.01$ for the Galactic diffuse emission and $0.97 \pm 0.01$ for the isotropic one. We also verified that there were no significant residuals from the whole time test-statistic map.
    
To obtain the light curve, we fitted the model in each time bin (3, 7, and 30 days), keeping the Galactic diffuse and the extragalactic isotropic normalisations fixed to their average values, and freeing the normalisation of the brightest sources within $8^\circ$ from the centre and with more than ten predicted photons. When Mrk\,501 was detected with a TS<25, we calculated 95\% flux upper limits. The distribution of the $\alpha$ parameter and the integrated flux for Mrk\,501 varied by less than $0.3\%$ and $1\%$, respectively, when we let the Galactic and extragalactic normalisation parameters free in the fit. If instead we freed more sources, $\alpha$ and the integrated flux might vary up to $3\%$ and $10\%$, respectively.

\subsection{TeV observations by FACT}
The First G-APD Cherenkov Telescope (FACT) is located in the Observatorio del Roque de los Muchachos on the island La Palma at 2.2\,km a.s.l.\ \citep{2013JInst...8P6008A}. Using the imaging air Cherenkov technique, it has been operational since October 2011.  With its 9.5\,m$^2$ segmented mirror, FACT detects gamma-rays with energies above a few hundred GeV by observing the Cherenkov light produced in extensive air showers induced by gamma and cosmic rays in the Earth's atmosphere. This approach allows for consistent long-term unbiased monitoring at TeV energies. Typically, no human interaction is needed during the night. The shift is performed fully remote and in an automatic way by software. In combination with an unbiased observing strategy, this allows for consistent long-term monitoring at TeV energies. This is facilitated by the excellent and stable performance of the camera using silicon-based photosensors (SiPM, also known as Geiger-mode avalance photo diods; G-APDs) and a feedback system, keeping the gain of the photosensors stable \citep{Biland_2014}. On the one hand, the stable and automatic operation maximises the data-taking efficiency, and with this, the duty cycle of the instrument. On the other hand, the usage of SiPMs also minimises the gaps in the light curves \citep{2017ICRC...35..609D} as these photosensors allow for observations with bright ambient light. While the telescope can operate during full-Moon conditions \citep{2013ICRC...33.1132K}, observations are typically interrupted for 4-5 days every month for safety reasons associated with the operational conditions at the observatory site.

Between December 14, 2012, and April 18, 2018, a total of 1953~hours of physics data from Mrk\,501 have been collected during 889~nights with up to 5.5~hours per night and an average nightly observation time of 2.2~hours. To ensure enough event statistics, nights with an observation time of less than 20 minutes were removed from the data sample. After this and after data quality selection (see below), 1344~hours from 633~nights remain. The FACT light curve is shown in Fig.~\ref{fig:LC} (uppermost panel). 

To derive the light curve, the data were processed using the modular analysis and reconstruction software\footnote{https://trac.fact-project.org/browser/trunk/Mars/} \citep[MARS, revision 19203;][]{2010apsp.conf..681B}. The feedback system enables stable and homogeneous gain that does not need to be recalibrated. After signal extraction, the
images were cleaned using a two-step process, as described in \cite{2021A&A...647A..88A}. 
Based on a principal component analysis of the remaining distribution, a set of parameters describing the shower image was obtained. A detailed description of the 
image reconstruction and background suppression cuts for light curve and spectral extraction is provided in \cite{2019ICRC...358..630D}.  
By cutting in the angular distance between the reconstructed source position and the real source position in the camera plane, $\theta$, (at $\theta^2 < 0.037$\,deg$^2$), the signal from the source was determined. To estimate the background, the same cut was applied for five off-source regions, each located 0.6\,deg from the camera centre (observations are typically carried out in wobble-mode \citep{1994APh.....2..137F} with a distance of 0.6\,deg from the camera centre). 

The excess rate was calculated by subtracting the scaled signal in these off-regions from the signal at the source position and dividing it by the effective exposure time.
To study and correct the effect of the zenith distance and trigger threshold (changing with ambient light conditions) \citep{2013arXiv1308.1516B}, the excess rate of the Crab nebula was used. While flares in the MeV/GeV range have been seen, similar flux changes have not been found at TeV energies. Therefore the Crab nebula was used as a standard candle at TeV energies. The dependences of the excess rate on observing conditions are similar to those of the cosmic-ray  rate described in \cite{2019APh...111...72B}, and details can be found in \cite{2019ICRC...358..630D}.

For the studied data sample, the corrections in zenith distance are smaller than 10\% for more than 97\% of the nights, and in trigger threshold, lower than 10\% for more than 75\% of the nights. 
The maximum correction in zenith distance is 44\%, and only two nights need a correction larger than 40\%. In threshold, the largest correction is 72\%, but 99\% of the nights 
have a correction smaller than 60\%.

To verify the effect of the different spectral slopes of Mrk\,501, the spectra of 35 time ranges between February 2013 and August 2018 determined using the Bayesian Block algorithm (see Sect.~\ref{sec:timing}) were extracted (see details in \cite{2015ICRC...34..707T} and background suppression cuts from \cite{2019ICRC...358..630D}) were fitted with a simple power law. Within the uncertainties, no obvious correlation was found between index and flux, so that the harder-when-brighter behaviour reported in \cite{2015A&A...576A.126A} is not confirmed for the TeV energy band. The distribution of indices yields an average spectral slope of
$2.96\pm0.26$,  
which is compatible with some previously published results of other telescopes, taking the different energy ranges and instrument systematics into account  \citep{2007ApJ...663..125A}.
Assuming different slopes from $2.7$ to $3.22$, the corresponding energy thresholds ($E_{Th} \sim 750\,GeV$ \citep{2017AIPC.1792e0019C}) and 
integral fluxes were determined to estimate the systematic error of 
a varying slope of the spectrum of Mrk\,501. This resulted in a systematic 
flux uncertainty lower than 7\%, which was added quadratically to the statistical uncertainties.

For the light curve of Mrk\,501, a data quality selection cut based on the cosmic-ray rate \citep{2017ICRC...35..779H} was applied to the data sample from December 14, 2012, until April 18, 2018 (MJD 56275 to 58226). For this, the artificial trigger rate $R750$  was calculated above a threshold of 750\,DAC-counts, which was found to be independent of the actual trigger threshold. The remaining dependence of $R750$ on the zenith distance was determined (as described in \citep{2017ICRC...35..612M} and \citep{2019APh...111...72B}). A corrected rate $R750_{\rm cor}$ was calculated based on this. Seasonal changes of the cosmic-ray rate due to variations in the Earth's atmosphere were taken into account by determining a reference value $R750_{\rm ref}$ for each moon period. The distribution of the ratio $R750_{\rm cor}$/$R750_{\rm ref}$ can be described with a Gaussian distribution for the good-quality data, while data with poo quality lies outside the Gaussian distribution. A cut was applied at the points at which the distribution of all data starts to deviate from the Gaussian distribution. Therefore, data with good quality were selected using a cut of $0.93 < R750_{\rm cor}/R750_{\rm ref} < 1.3$ \citep{2021A&A...647A..88A}. This resulted in a total of 1344~hours of good-quality data from Mrk\,501 during the discussed time period, distributed over 633 nights, as shown in Fig.~\ref{fig:LC} (uppermost panel).

\section{Timing analysis\label{sec:timing}}

In this section, we investigate the variability of  Mrk\,501 band by band using fractional variability and auto-correlation functions.
We also use Bayesian Block analysis to identify individual flares and distinguish different source states. Temporal relations between different spectral bands are probed using cross-correlation functions and convolution techniques.

\subsection{Variability\label{sec:fvar}}

The variability in light curves can be compared and quantified by the unit-less fractional variability 
$F_{var} = \sqrt{(S^2 - \langle\sigma^2_{err}\rangle)/\langle{x}\rangle^2}$ \citep{Vaughan_2003MNRAS.345.1271V},
where S is the standard deviation of the light curve, $\langle\sigma^2_{err}\rangle$ is the mean squared flux error, and $\langle{x}^2\rangle$ is the square of the average flux.
Uncertainties of the fractional variability were estimated using the prescription by \cite{Poutanen_2008MNRAS.389.1427P} and \cite{Vaughan_2003MNRAS.345.1271V}.

\begin{figure}
  \centering
      \includegraphics[width=\columnwidth]{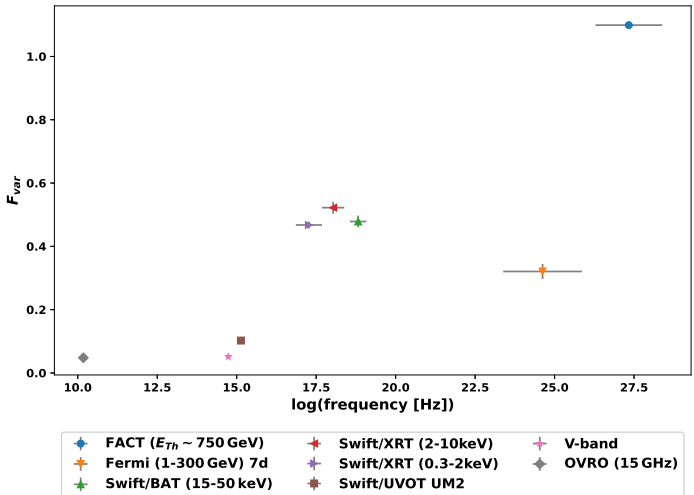}
  \caption{Fractional variability $F_{var}$ as a function of frequency. X-axis error bars indicate the energy band of the instrument. Y-axis error bars denote the uncertainty of $F_{var}$ , which for some instruments are smaller than the marker.}
  \label{fig:fvar_all}
\end{figure}

The dependence of the fractional variability on frequency, calculated for the light curves spanning about 5.5 years, is depicted in Fig.~\ref{fig:fvar_all}. 
The fractional variability is lowest in the radio and optical, increases up to the X-rays, drops in the GeV, and again increases up to the TeV, where the highest variability is found. These two maxima indicate that the falling edges of the two SED components are more variable than the rising ones and that the component peaking in the TeV varies more than the component peaking in the X-rays.
Similar behaviour was observed in Mrk\,421 \citep{2015A&A...576A.126A, aleksic_2015A&A...578A..22A,2021A&A...647A..88A}. 
Some previous studies of Mrk\,501 focusing on the study of large flares found a fractional variability that monotonically increased with frequency \citep{2016A&A...594A..76A, ahnen_2017A&A...603A..31A}. Different binning, cadence, and coverage in the different wavebands may play a role in this discrepancy \citep{2019Galax...7...62S}.

We calculated the structure functions \citep[SF,][]{1985ApJ...296...46S} of the various light curves and found in the X-rays an average SF slope of 0.13 and first break at around 20\,days. In the TeV range, the average slope of the SF is 0.09. In the radio, the SF slope is close to zero, suggesting either white noise rather than well-structured flares or numerous overlapping flares.

\subsection{Light-curve correlations\label{sec:crosscor}}

We correlated the light curves either with themselves (auto-correlation) or with other light curves using the discrete correlation function \citep[DCF,][]{1988ApJ...333..646E}, which properly takes the irregular sampling of the  data into account. All the correlations were calculated by filtering out the data of low significance (<$2 \sigma$) and additionally discarding all upper limits from the {\it Fermi}-LAT light curve, as is often done in the literature \citep[e.g.][]{2011ApJ...738...25A, 2018A&A...620A.181A}. The uncertainties were calculated with the DCF algorithm.

\begin{figure}[h!]
  \centering
      \includegraphics[width=0.9\columnwidth]{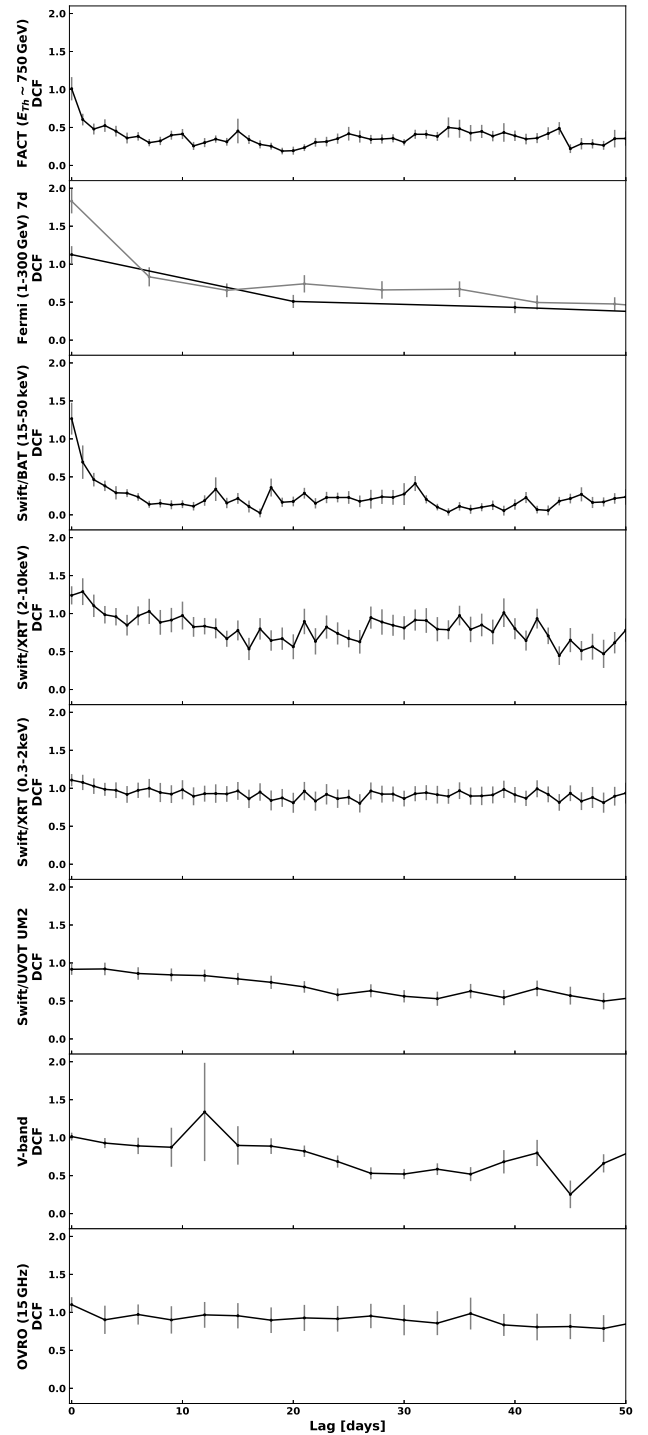}
  \caption{Light-curve auto-correlations. From top to bottom: FACT, {\it {\it Swift}}/BAT, {\it Swift}/XRT (2-10\,keV), {\it Swift}/XRT (0.3-2\,keV), {\it Fermi}-LAT, {\it Swift}/UVOT UM2, V-band, and radio. Grey error bars denote 1$\sigma$ DCF uncertainties. A one-day binning was used, except for {\it Fermi}-LAT (20- and 7-day lag binning, black and grey lines, respectively) and for the UV, optical, and radio data (3 days).}
  \label{fig:acf_all}
\end{figure}

A time bin of one day was used in the discrete auto-correlation function (DACF) for most data, except in the UV, optical, and radio (three days) and in the GeV.
As the {\it Fermi}-LAT flux uncertainties are correlated because the same sky model was used for the different time bins, we had to use a time bin of 20 days \citep[see the discussion in][]{1988ApJ...333..646E}. Shorter lags would result in a correlation above unity (see the grey curve in Fig.~\ref{fig:acf_all}). For the TeV and X-rays cross-correlations, we used a one-day lag binning, while for the radio and optical correlations with GeV, we used three days.

The auto-correlations (Fig.~\ref{fig:acf_all}) indicate the presence of flares developing on a timescale of a few days mostly in the TeV and hard X-rays and of variations occurring on timescales
of many weeks. As the shortest variability timescale of Mrk 501 in the X-rays \citep[$\sim$one day,][]{2001ApJ...563..569T} is shorter than our time bins, the correlation peaks characterise flaring patterns longer than the binning period. 

We tested the cross-correlations between all light curves and discuss the significant ones (TeV/X-rays, UV/V-band, and GeV/radio) in Sect. \ref{sec:tevx}, \ref{sec:gevx}, and  \ref{sec:crosscor}. 

\begin{figure}[b]
  \centering
       \includegraphics[width=\columnwidth]{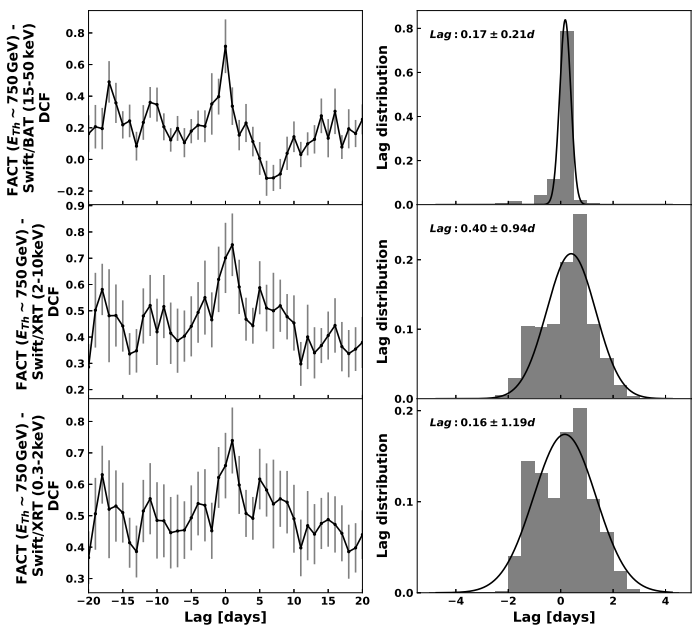}
       \caption{DCF cross-correlations of light curves (from top to bottom panel): FACT with {\it Swift}/BAT, {\it Swift}/XRT (2-10\,keV), and {\it Swift}/XRT (0.3-2\,keV). One-day binning was used. Left: DCF values as a function of lag. Grey error bars are 1$\sigma$ uncertainties. Right: Lag distributions derived from FR/RSS simulations (more details are provided in Sect.~\ref{sec:tevx}). A Gaussian fit (black lines) was used to derive the lag indicated in the plots.}
  \label{fig:dcf_all}
\end{figure}

\begin{figure}
  \centering
       \includegraphics[width=\columnwidth]{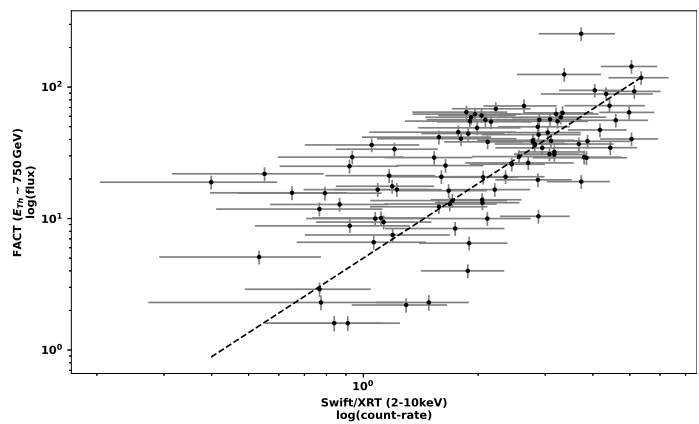}
       \caption{TeV (FACT) flux and X-ray ({\it Swift}/XRT 2-10\,keV) count-rate measured within 24 hours (slope $1.89\pm0.15$).}
  \label{fig:xray_tev_flux}
\end{figure}

\subsection{TeV / X-ray correlation\label{sec:tevx}}

A strong correlation reaching 0.7 at 0 lag is found between the TeV (FACT) and X-ray ({\it Swift}/XRT or {\it Swift}/BAT) light curves (Fig.~\ref{fig:dcf_all}). The correlation peak is wider for {\it Swift}/XRT because of the larger (1-2 days) relative time distances between the FACT and {\it Swift}/XRT observations. In case of {\it Swift}/BAT, observations are coincident (within a day) with the FACT light curve.

The lag probability distributions reported in Fig.~\ref{fig:dcf_all} are obtained from Monte Carlo simulations \citep[as in][]{Peterson_1998PASP..110..660P}. We generated $10^4$ subsets for each pair of light curves using flux randomisation (FR) and random subset selection (RSS) and calculated the resulting DCFs to obtain a representative time lag distributions using a centroid threshold of 80\% of the DCF maximum \citep{2004ApJ...613..682P}. The lag uncertainty corresponds to the standard deviation of the distribution of the lag values obtained for the random subsets.

The best estimate of the TeV/X-ray lag of $(0.17\pm0.2)$ days is obtained by correlating the TeV and hard X-ray {\it Swift}/BAT light curves. Summing the time-lag distributions obtained from soft and hard X-rays increases the uncertainty to $(0.3\pm0.4)$ days.
This TeV/X-ray correlation in Mrk 501 was already reported using shorter and sparser data sets \citep{2017ApJ...841..123P,2018A&A...620A.181A}. 

As noted earlier, the TeV variability is stronger than in the X-rays, with a flux correlation slope (Fig.~\ref{fig:xray_tev_flux}) increasing from $1.3\pm0.1$ (with reference to the variability observed by {\it Swift}/XRT 0.3-2\,keV) to $1.89\pm0.15$ if {\it Swift}/XRT 2-10\,keV is used. The dispersion of the data points in Fig.~\ref{fig:xray_tev_flux} around the main correlation, obtained using the orthogonal distance regression method, is compatible with intra-day variability in these bands or/and by spectral variations between flares. To build the correlation plots (Fig.~\ref{fig:xray_tev_flux} and Fig.~\ref{fig:spectral_properties}), only coincident FACT and {\it Swift}/XRT observations were considered with measurements in time separation of less than 12 hours.

Bayesian Block algorithm was used to identify individual flares in the TeV and X-rays bands. The algorithm properly takes the significance of the data points into account, so that we applied it to the unfiltered data (see Sect.~\ref{sec:data} and Fig.~\ref{fig:LC}, where the uncleaned light curves are shown). The Bayesian Block algorithm was tuned for a false-positive probability of 1\% \citep{2013ApJ...764..167S}. A flare is defined as a significant change in flux (3$\sigma$) with a duration of at least one binning point. Thirty-seven TeV flares were identified. To enhance the detection of X-ray flares, we compiled a list of individual flares detected by {\it Swift}/BAT and {\it Swift}/XRT separately. Unfortunately, as the X-ray data were still often either too sparse or too noisy, we finally considered only 15 TeV flares (56330 -- 56355, 56462 -- 56465, 56467 -- 56472, 56491 -- 56498, 56536 -- 56540, 56541 -- 56545, 56815 -- 56818, 56831 -- 56833, 56856 -- 56862, 56874 -- 56883, 56887 -- 56893, 56936 -- 56944, 57156 -- 57171, 57230 -- 57235, and 57483 -- 57489), with simultaneous coverage in X-rays. All of them coincide with individual X-ray flares or with periods with some level of X-ray flaring activity. TeV and X-ray flares have a similar duration. Most flares last less than seven days, confirming the cross-correlation results. 

\subsection{GeV / soft X-ray correlation\label{sec:gevx}}

\begin{figure}[h]
  \centering
      \includegraphics[width=\columnwidth]{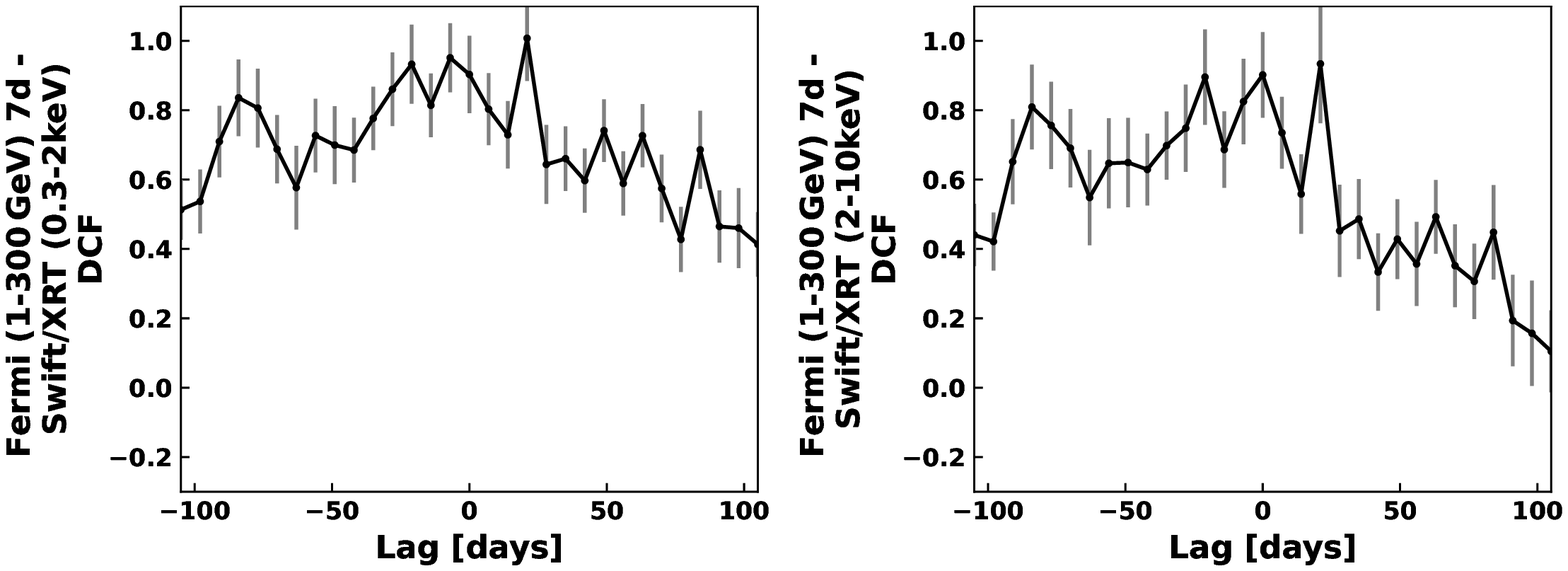}
      \includegraphics[width=\columnwidth]{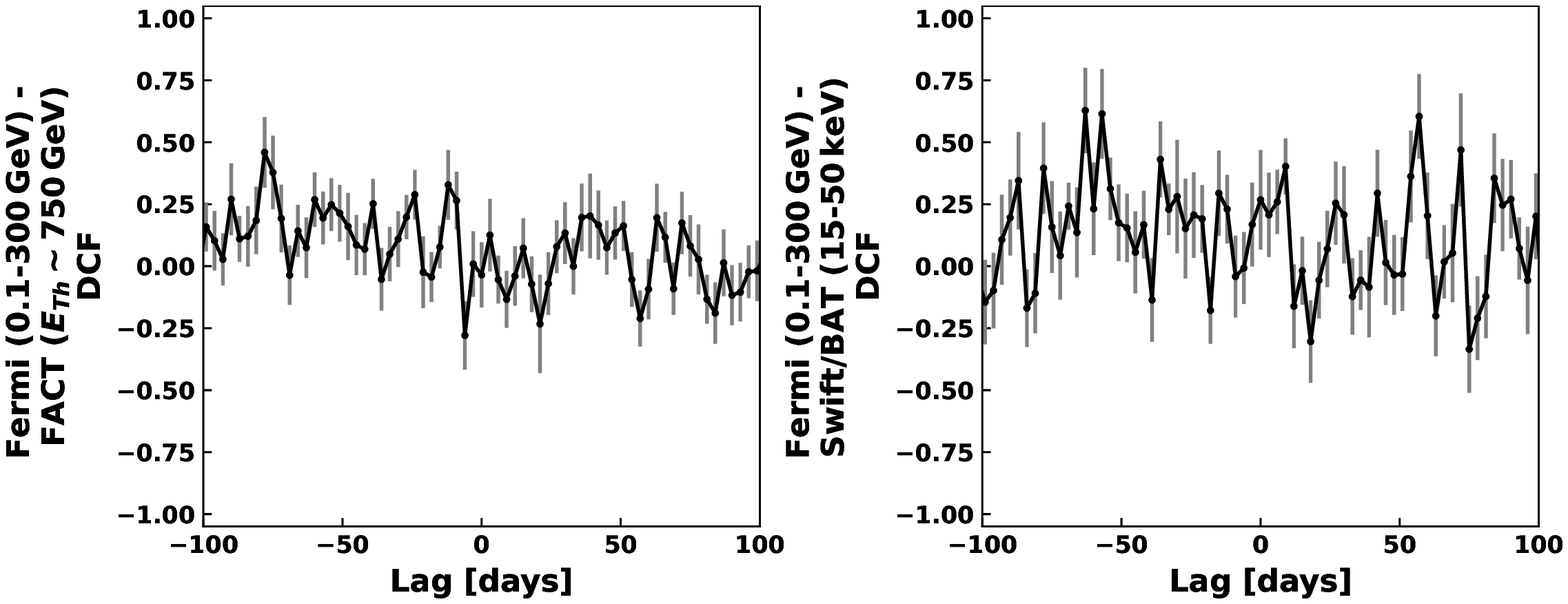}
  \caption{Cross-correlation of the GeV {\it Fermi}-LAT light curve with the data from {\it Swift}/XRT 0.3-2\,keV (top right), {\it Swift}/XRT 2-10\,keV (top left), FACT (bottom right), and {\it Swift}/BAT (bottom left). The time resolution is seven days. The grey error bars denote 1$\sigma$ uncertainties.}
  \label{fig:cc_fermi_fact_bat}
\end{figure}

The GeV and soft X-ray variations appear strongly correlated on long timescales, as depicted in Fig.~\ref{fig:cc_fermi_fact_bat} (top panel). The GeV light curve is not correlated with the hard X-ray ({\it Swift}/BAT) or TeV light curves, however. This indicates that the GeV variability is not simultaneous to the flares lasting a few days, but is dominated by the long-lasting variations. This behaviour may be biased by the low sensitivity in the GeV band, and should only be considered along with other temporal data.

\subsection{Correlations at/with longer wavelengths\label{sec:crosscor}}

\begin{figure}
  \centering
      \includegraphics[width=\columnwidth]{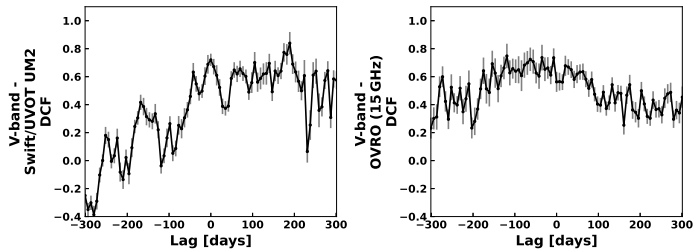}
  \caption{Multi-wavelength correlations for different bands. Left: Cross-correlation of the V-band and {\it Swift}/UVOT light curves. Right: Cross-correlation V-band and radio light curves. Grey error bars denote 1$\sigma$ uncertainties.}
  \label{fig:um2_radio_vband}
\end{figure}

We find a strong (reaching 0.85) but broad correlation between the optical and ultraviolet variations (Fig.~\ref{fig:um2_radio_vband}), and both are broadly correlated with the radio light curve. Lags between the bands cannot be reliably estimated from the cross-correlations alone (Fig.~\ref{fig:acf_all}).

The correlation between the GeV and radio light curves also appears strong (reaching 0.8), with radio lagging behind the GeV by 170-250 days. 
GeV-radio delays have also been reported in various sources by \cite{max-moerbeck_2014MNRAS.445..428M}, and were interpreted as shocks propagating in the jet. 

\begin{figure}
  \centering
      \includegraphics[width=\columnwidth]{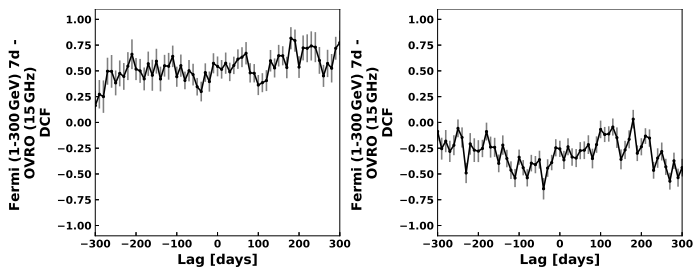}
  \caption{Multi-wavelength correlations for different bands. Left: Cross-correlation of the {\it Fermi}-LAT and radio light curves in the time range of [56800, 58226] MJD. Right: Cross-correlation of the {\it Fermi}-LAT and radio light curves in the time range of [54693, 56800] MJD using data from \citep{max-moerbeck_2014MNRAS.445..428M}. Grey error bars denote 1$\sigma$ uncertainties.}
  \label{fig:fermi_radio}
\end{figure}

The radio-GeV behaviour of Mrk\,501 appears very different before the period considered in this study, however. Before MJD 56800, no correlation could be found (Fig.~\ref{fig:fermi_radio}) using the data from \cite{max-moerbeck_2014MNRAS.445..428M}. All the other flux correlations   derived above for the full data set do not change notably when data either before or after MJD 56800 are selected. The source may experience an internal change (misalignment due to the precession of a mini-jet, emission region properties change, etc.), which yields different behaviour and inter-band connection before and after 56800 MJD. It is important to note that this change is not related to the source flaring activity state, and there is no such change after 57600 MJD, for example, when Mrk\,501 reaches a quiescent state in most bands (Fig.~\ref{fig:LC}). 

\subsection{Spectral variations\label{sec:specvar}}

\begin{figure}
  \centering
      \includegraphics[width=\columnwidth]{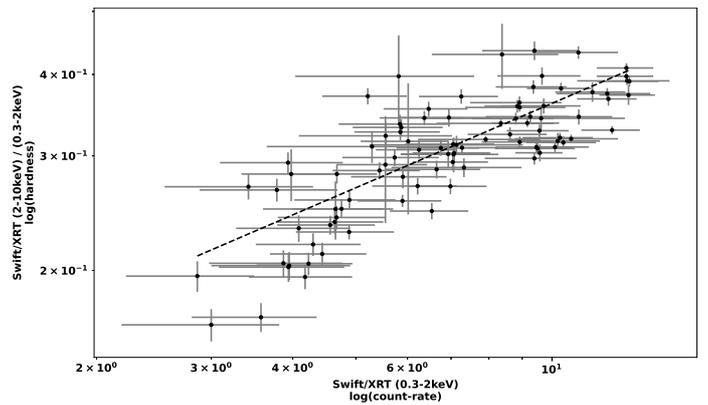}
  \caption{Ratio of {\it Swift}/XRT hard-to-soft vs {\it Swift}/XRT 0.3-2\,keV count rate. The slope is $0.43\pm0.04$.
  }
  \label{fig:spectral_properties_xrt}
\end{figure}

\begin{figure}
  \centering
      \includegraphics[width=\columnwidth]{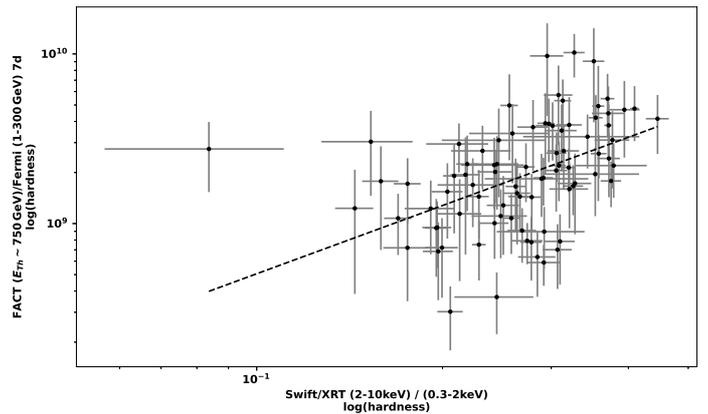}
  \caption{Ratio of TeV to GeV fluxes vs the ratio of {\it Swift}/XRT hard-to-soft flux. The slope is $1.4\pm0.3$.}
  \label{fig:spectral_properties}
\end{figure}

The long-term duration of our campaign allows us to study the connection between spectral components. In the X-rays, we find a clear indication for a long-term harder-when-brighter behaviour (Fig.~\ref{fig:spectral_properties_xrt}). The variability amplitude increases at harder X-rays right of the low-energy hump maximum in the SED (Fig.~ \ref{fig:spectral_properties_xrt}). Along with Fig.~\ref{fig:xray_tev_flux}, this indicates a higher variability beyond the cut-off energies of both spectral components. The colour-colour diagram, showing the ratios of the TeV and GeV fluxes and of the hard and soft X-rays fluxes, respectively (Fig.~\ref{fig:spectral_properties}), confirms the relation between the two spectral components and that the variability amplitude of the high-energy component is higher than that of the low-energy component.

\section{Discussion\label{sec:discussion}}

\subsection{Summary of results}

We studied the broadband variability of Mrk\,501 from the end of 2012 to the middle of 2018. Data from eight instruments were considered. During this period, the source experienced numerous flaring periods in the TeV. The highest flux was measured in 2014. 

The variability was detected in all wave bands. The fractional variability is lowest in the radio ($F_{var} \sim 0.05$) and highest in the TeV ($F_{var} \sim 1.1$), and it monotonically increases from the radio to the X-rays and  from the GeVs to the TeVs. A similar fractional variability pattern was reported for Mrk\,421 on monthly \citep{2015A&A...576A.126A} and yearly \citep{2016A&A...593A..91A} timescales. In the context of the one-zone synchrotron-self-Compton (SSC) model, this variability pattern indicates that the electron spectrum is more variable at higher energies. For the same period of observations, Mrk\,501 is less variable than another nearby and bright HBL blazar Mrk\,421 \citep{2021A&A...647A..88A}.

To determine correlations between the spectral bands, we calculated the discrete-correlation functions for pairs of light curves and found that the X-ray and TeV emission is well correlated with a lag $<0.4$  days $(1\sigma)$. All the flares detected in the TeV band were coincident with X-ray flares. 

The fractional variability and the correlated TeV and X-ray emission are likely produced by a synchronous change of the spectral shape of the low- (X-rays) and high-energy (TeV) components. 
A zero-lag correlation between the X-ray and TeV light curves suggests that the variations in the two bands are driven by a common physical process operating in the same emission region. We compare the observational constraints to different models in the next subsections, assuming that the flares are related to particles accelerated in shocks.

The radio and GeV emissions were correlated during the majority of the campaign (after about 56800 MJD), but not in older data (Fig.~\ref{fig:fermi_radio}, left and right), indicating that the long-term radio emission showing the least variability originates from several components. 
The VLBA high-resolution radio images at 43\,GHz indeed reveal an off-axis jet component \citep{koyama_vlbi_aa_2016_586} that may explain that the GeV and radio are not always correlated \citep[see also ][]{2008A&A...488..905G,2016A&A...586A.113K}.  
In July 2014, the radio-optical emission of Mrk\,501 was also interpreted as a separate spectral component with little contribution in the GeV \citep{2020A&A...637A..86M}.

\subsection{Synchrotron self- or external Compton emission\label{sec:synchrotron}}

In the quiescent state, several multi-wavelength campaigns of Mrk\,501 reported that its spectral energy distribution was compatible with a one-zone SSC model  with a Doppler beaming factor $\delta\approx 12-21$. \citep{2020A&A...637A..86M, 2018PASJ...70...45L,2011ApJ...727..129A,2007ApJ...669..862A}. 
Additional spectral features detected during flares \citep{2020A&A...637A..86M, 2018PASJ...70...45L, ahnen_2017A&A...603A..31A} led to higher $\delta\approx 10-50$.
The wide range of $\delta$ values can be explained by a deceleration of the jet \citep{2003NewAR..47..653G} or by a radial structure with an inner fast spine and a slower outer layer where the radio emission is produced \citep{2005A&A...432..401G}. In both cases, the expected variability increases with frequency and should be lowest in the radio.

One of the most significant constraints on the size of the emitting region can be derived from the variability timescale. As expected from geometrical properties of a homogeneous source of incoherent, isotropic comoving-frame emission with a comoving size $R$, the minimum expected variability timescale can be written as $t_{var_{min}} = (1 + z)R/(\delta c)$ (where $z$ is the source redshift). The shortest doubling timescale observed in Mrk\,501 is 2 minutes \citep{2011ApJ...727..129A}, and the delay between various VHE bands is 4 minutes (an indication for progressive electron acceleration) indicate a gamma-ray source size $<10^{12}$\,km.

Variability can also be constrained by the synchrotron cooling timescale. Synchrotron cooling time in the observer's frame of reference can be written as $t_{cool,\,e} \approx 15.86\times10^{11} \left((1+z)/\delta\right)^{1/2} \,(B/1{\rm G})^{-3/2} (\nu/1{\rm Hz})^{-1/2}\,{\rm seconds}$ \citep{2019ApJ...884..125Z}, where $\nu$ is the emission frequency. Using typical values of $B$ from $0.01$ to $0.31$ G, and $\delta$ from 12 to 25 \citep{2011ApJ...727..129A, ahnen_2017A&A...603A..31A}, we can find that the synchrotron cooling times in the observer frame become shorter than 9 minutes, 2 hours, 7 hours, and 176 days for 50\,keV, 0.3\,keV, in the V-band, and in the radio, respectively. These timescales are much shorter than those shown in the auto-correlations.
The duration of the flares should therefore be driven by physical changes of the emitting source(s) occurring on dynamical time scales.

The reported delay $<0.4$ days between the TeV and X-rays fluxes agrees with the self-Compton or external Compton frameworks as electrons cool rapidly $(<0.5\, {\rm hour})$ at these energies. The observed connection between the X-ray and TeV emission (Fig.~\ref{fig:xray_tev_flux}) also supports the SSC model scenario of inverse Compton production of VHE gamma-rays by the same population of electrons. The relation between the X-ray and gamma-ray spectral breaks (Fig.~\ref{fig:spectral_properties}) finally indicates that the cut-off energies of both spectral components are also related. 

Assuming self-Compton emission for the TeV $\gamma$-rays \citep[Mrk 501 is always in the Thomson regime in the TeV, e.g. ][]{2012ApJ...749...63M}, the scenario with the stronger magnetic field produces a minute-variability timescale, which is close to the observations ( \cite{2011ApJ...727..129A}). Taking the compactness of the VHE source region into account, high Doppler factors ($\delta_{flaring} > 40$, $\delta_{low state} > 30$) \citep{2017AIPC.1792e0019C} are needed to avoid $\gamma$-ray absorption. An explanation for the fast TeV variability may also come from MHD turbulent flows or phase transitions \citep{1994ApJ...423..172K}, instabilities of hydrodynamic flows \citep{2012MNRAS.423.1707S}, or jet or mini-jet beaming turbulences on different scales \citep{1998A&A...338..399C, 2010MNRAS.402.1649G}.

In the one-zone synchrotron regime, the energy cutoff and the flux normalisation are controlled by $B$, $\delta,$ and the particle spectrum. Variations in these parameters can lead to independent GeV and TeV variability (as indicated by the observations, see Fig.~\ref{fig:cc_fermi_fact_bat}). In addition, in the shock-in-jet model, the regions emitting  X-rays and TeVs are closer to the black hole than those responsible for the bulk of the optical-radio and the GeV emissions. In the SSC frame, the observed correlations between TeV and X-rays (fluxes and spectral slopes), radio, and GeV fluxes, and the lack of correlations between the TeVs/X-rays and GeVs would require that some of the above physical parameter(s) is/are changing along the jet.

\subsection {Hadronic and lepto-hadronic radiation scenarios}
\label{subsect:alt_models}

In addition to pure leptonic scenarios, in which all radiation is produced by relativistic electrons, purely hadronic models \citep{2001APh....15..121M} and lepto-hardronic models were also proposed \citep{2013MNRAS.434.2684M,2016APh....80..115P,2001ICRC....3.1153M,2001APh....15..121M,2020MNRAS.491.5268G,1993A&A...269...67M}. 

In lepto-hadronic models, the low-energy SED component is still dominated by the synchrotron emission of primary accelerated electrons. However, the high-energy SED component is dominated by proton synchrotron and/or pion-decay-induced pair cascades. 
The correlation between the TeV and the X-ray flares indicates that electrons and protons are accelerated in the same region.
As the proton acceleration time $t_{acc}=20\,\xi\,\gamma\,m_p c/3\,eB \geq 10\,\xi\ {\rm days}$ \citep[where $\xi\geq 1$ is the mean free path in units of Larmor radius,][]{2000ApJ...536..299K, 1996ApJ...463..555I} is much longer than the delay observed between the TeV and X-ray light curves, the driving mechanism of the variability should be linked to some dynamical processes (e.g. changing orientation or beaming of the relativistically moving emission region) in combination with propagation effects. However, the absence of a correlation between the GeV and TeV bands on the same timescale as observed for the TeV/X-ray correlation suggests that the variability of the hadronic component is not driven by the process behind the variability of the leptonic component.

Hadronic emission models propose proton synchrotron to be responsible for the low-energy SED component and pion-induced cascades for the high-energy component (Mastichiadis et al.2013). These models require strong magnetic fields (>10 G) such that the Larmor radius of the protons remains smaller than the jet radius \citep{2011IAUS..275...59S}. In these conditions and assuming that the X-ray emission is from proton synchrotron, the expected cooling timescale of tens of hours \citep{2019ApJ...884..125Z} is also inconsistent with observations. Variability on shorter timescales may again possibly be driven by dynamical processes, which affect the whole population of relativistic particles, but as above, this is not consistent with the absence of a GeV/TeV correlation and with radio lagging behind the GeV by about 200 days.


The mechanisms responsible for the broadband emission of blazar jets, accelerating electrons should in principle also accelerate protons and heavier nuclei possibly related to the observed ultra high-energy cosmic rays (UHECRs). When these particles interact with the surrounding photons, $\gamma$-rays and neutrinos may be produced \citep{2018MNRAS.480..192P,2018ApJ...863L..10A,1993A&A...269...67M}. Muons and pions created in these processes will also be affected by the proton acceleration time, so that timescales observed in the TeV cannot be reproduced with this mechanism \citep{2017A&A...602A..25Z}. In addition,
Pion photoproduction is generally inefficient because protons mostly lose energy through synchrotron radiation \cite[but see][]{1998PhRvD..58l3005R}.

\subsection{Flare timing}

\begin{figure}
  \centering
      \includegraphics[width=\columnwidth]{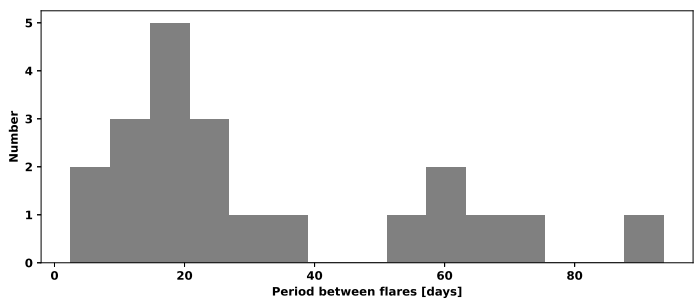}
  \caption{ Period between peaks of 37 identified TeV flares, 15 of which have good X-ray coverage and are listed in Sect. \ref{sec:tevx}. }
  \label{fig:flares_periods}
\end{figure}

The mass of the supermassive black hole (SMBH) of Mrk\,501 is estimated as $0.9-3.4\times10^{9}M_{\odot}$ \citep{Barth_2002}. 
The distribution of the time delays between maximums of identified TeV flares peaks between 5 and 25 days (Fig.~\ref{fig:flares_periods}) (undetected flares could mimic longer delays). These delays correspond to $\sim10^{4}\ R_G/c\sim 17$ days, which could be expected if the flares were related to Lense-Thirring \citep{1918PhyZ...19...33T} precession of a misaligned accretion disk \citep{1975ApJ...195L..65B,10.1093/mnrasl/slx174}. The inter-flare period originating from such a precession is shortened in the source rest frame by a factor $\Gamma$, then, for the observer, it becomes shortened by a factor $\Gamma^{-1}$, which compensate for each other.

\section{Conclusions\label{sec:conclusions}}

The analysis of about 5.5 years of light curves from eight instruments yields the following main observational results:
\begin{enumerate}
\item The strongest variations were found in TeV and X-rays. The TeV and X-ray fluxes measured simultaneously (within 24~hours) are correlated, as are the X-ray and gamma-ray spectral breaks, as expected from SSC models. The lag between the TeV and X-ray variations could be estimated as $<0.4$ days $(1\sigma)$.

\item The characteristic time interval between TeV flares is comparable with the expectation if these flares are triggered by a Lense-Thirring precession of the accretion disk around the SMBH.
\end{enumerate}

The observed variability of Mrk\,501 was compared with the predictions of leptonic, lepto-hadronic, and hadronic models. We found that purely hadronic models are incompatible with the observations due to extremely long synchrotron cooling time in the radio and relatively short GeV-radio delay. The lepto-hadronic models are also incompatible with observations because they fail to reproduce simultaneously the short lag observed between X-rays and TeV and the absence of correlation between the GeV and TeV. Electron synchrotron self- or external Compton processes match the observations of Mrk\,501 in general, although individual flares may require more complex models, including the introduction of a second emission zone \citep{abdo_2011ApJ...727..129A, 2020ApJ...901..132S,2018A&A...620A.181A}.

\begin{acknowledgements} 
We acknowledge the important contributions from ETH Zurich (grants ETH-10.08-2, ETH-27.12-1) and by the Swiss SNF, and the German BMBF (Verbundforschung Astro- und Astroteilchenphysik) and HAP (Helmoltz Alliance for Astroparticle Physics) to the FACT telescope project. 
Part of the reported work was supported by Deutsche Forschungsgemeinschaft (DFG) within the Collaborative Research Center SFB 876 "Providing Information by 
Resource-Constrained Analysis", project C3. We would like to express our gratitude to E. Lorenz, D. Renker and G. Viertel for their invaluable contribution to the FACT project on its early stages.
We would like also to thank the Instituto de Astrof\'{\i}sica de Canarias (IEC) for hosting the FACT telescope over the years and supporting its operations at the Observatorio del Roque de los Muchachos in La Palma. We thank Max-Planck-Institut f\"ur Physik for the provided HEGRA CT3 mount, which was refurbished and reused for FACT. We express our sincere gratitude to all MAGIC collaboration for taking care of FACT and help during its remote operations. This research used public data from the Bok Telescope on Kitt Peak and the 1.54 m Kuiper Telescope on Mt. Bigelow \citep{2009arXiv0912.3621S}, \textit{Fermi}-LAT \citep{2009arXiv0912.3621S} and {\it Swift} \citep{2004NewAR..48..431G}. This research has made use of data from the \textit{OVRO} 40-m monitoring program (Richards, J. L. et al. 2011, ApJS, 194, 29), supported by private funding from the California Insitute of Technology and the Max Planck Institute for Radio Astronomy, and by NASA grants NNX08AW31G, NNX11A043G, and NNX14AQ89G and NSF grants AST-0808050 and AST- 1109911. Data from the \textit{OVRO} 40-m telescope used for this study, was publicly available when we started the study, though as of the paper submission, the policy has changed and it is only available upon request.
\\
{\it Author contributions.} V.S. and R.W. performed the multi-wavelength light curves analysis and wrote the paper. M.B. performed the {\it Fermi}-LAT analysis, D.D. performed the FACT analysis, V.S. performed the {\it Swift}/UVOT analysis and R.W. performed the {\it Swift}/BAT analysis. A.A.E., A.B., T.B., D.D., M.B. \& K.M. provided comments and discussed the results. All authors contributed to the construction and/or operations of the FACT telescope.
\end{acknowledgements}

\bibliographystyle{aa}
\bibliography{9_references}
\end{document}